# Chromium Oxide Formation on Nanosecond and Femtosecond Laser Irradiated Thin Chromium Films

L. Kotsedi[1-3], V. Furlan[4], V. Bharadwaj[5,*], K. Kaviyarasu[1-2], B. Sotillo[5], C.B. Mtshali[1-2], N. Matinise[1-2], A. G. Demir[4] B. Previtali[4], R. Ramponi[5], S.M. Eaton[5], M. Maaza[1-2]

[1]*UNESCO-UNISA Africa Chair in Nanosciences-Nanotechnology, College of Graduate Studies, University of South Africa, Muckleneuk ridge, PO Box 392, Pretoria-South Africa.*

[2]*Nanosciences African Network (NANOAFNET), iThemba LABS-National Research Foundation, 1 Old Faure road, Somerset West 7129, PO Box 722, Somerset West, Western Cape Province, South Africa.*

[3]*University of the Western Cape, Physics Department, Robert Sobukwe Road, Bellville Cape Town, 7535*

[4]*Department of Mechanical Engineering, Politecnico di Milano, Via La Masa, 1, 20156 Milano, Italy.*

[5]*Institute for Photonics and Nanotechnologies (IFN) – CNR, Piazza Leonardo Da Vinci, 32, 20133 Milano, Italy.*

*corresponding author: vibhavbharadwaj@gmail.com

**Abstract**:

Thin coatings of $Cr_2O_3$ have been used for numerous applications. Selective oxidation of chromium will be beneficial for integrated device fabrications. Thin coatings of pure chromium were vacuum deposited on a glass substrate using hot electrons from tungsten filament. These coatings were then treated with nanosecond and femtosecond laser in ambient conditions. The microstructure, morphology and the color of the coatings treated with laser sources were modified and there was a formation of an oxide layer due to the heat dissipation on the chromium coating from the energetic photons. High-resolution scanning electron microscope studies showed the morphological evolution, which is directly correlated with the laser fluence of both the nanosecond and femtosecond lasers. This morphological evolution was accompanied by the microstructural change as observed from the X-ray diffraction patterns. The chromaticity response of the coating was studied by UV-Vis spectroscopy and the response of the coating in the visible region evolved with the laser fluences. The divergence in chromaticity of these two laser treatments, is due to the difference in morphology as the result of the varied pulse duration. It could be concluded that the morphology had effect on the chromaticity of the films. Futhermore, Rutherford backscattering depth profiling of the laser treated coatings revealed the diffusion of oxygen atoms in the coating as a result of laser treatment fluence. We have analyzed both the optical and material properties of the laser induced oxidation and demonstrated laser writing as a promising tool to selectively oxidize Chromium for integrated device applications.

## 1. Introduction

Chromium is a [Ar] $3d^5 4s^1$ transition metal that can have amphoteric behavior, having nine oxidation states. The oxides of chromium can have varying stoichiometry depending on the method/protocol of preparation. These oxides of chromium can be synthesized in aqueous conditions using wet chemical methods [1], in powder form, when the powder is oxidized in air or oxygen ambient [2], or in a solid metallic state, when a slab of chromium metal is annealed in isothermal conditions [3]. Thin coatings of chromium deposited on any support of choice using physical or chemical deposition techniques, can also be oxidized using non-isothermal and non-equilibrium experimental conditions [4]. $Cr_2O_3$ is reported to be the hardest oxide state of chromium offering good wear resistance and low coefficient of friction [5]. Thin coatings of $Cr_2O_3$ have been used for applications as lubricant coatings in gas bearings [6], as the absorbing material in solar cells [7], as protections for magnetic data recording devices [8] and as shields in flexible solar cells [9]. Selective oxidation of chromium will be beneficial for integrated devices for the applications mentioned.

Lasers have been applied for surface structuring and oxidation of thin films [10]. Direct laser writing using pulsed lasers offers a possibility for spatially selective oxidation of chromium which can be tailored to a given application. Focused pulsed laser beams are selectively absorbed in the focal spot on thin Chromium layers deposited on a substrate, leading to material modifications due to heating and eventually oxidation from the moisture in the air [11-12]. By translating the sample with respect to the focal spot, oxidation can be mediated in various 2D geometries on the material. A schematic depicting direct laser writing on Chromium thin films is shown in Figure 1.

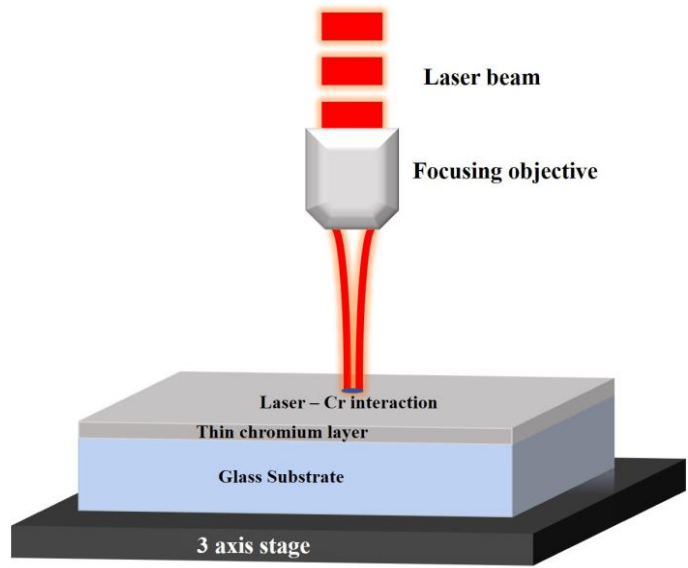

**Figure 1**. Schematic representing direct laser processing on Chromium thin films.

Numerous laser sources with varying fundamental wavelengths and pulse durations have been used to achieve laser heating of chromium thin coatings. Kotsedi *et al.* [13] showed the use of 1064 nm fundamental wavelength femtosecond laser to heat thin coatings of chromium, and from the X-ray diffraction studies, the $Cr_2O_3$ phase was observed, confirming the oxidation of the chromium thin coatings. Lian *et al.* [14] has employed 1062 nm fundamental wavelength laser source with ms - pulse durations, to observe the oxidation of the chromium coating. Their studies were performed to understand the oxidation rate dynamics. Veiko *et al.* [15] successfully oxidized chromium thin coatings using a picosecond laser with a fundamental wavelength of 1064 nm, confirming the formation of an oxide layer by Raman spectroscopic studies. They also report the formation of $CrO_2$ state, which is difficult to detect using x-ray diffraction technique. However, a morphological study and depth profiling of the oxidized coating has not been performed until now.

The laser-matter interactions during direct laser writing on thin films leading to heating, ablation or crystallization [16-17] has a direct effect on the morphological, microstructural or crystallographic phase evolutions. These phenomena are strongly dependent on the writing parameters, such as the wavelength, pulse duration, repetition rate and laser fluence. Here, we have applied industrially relevant laser sources to study a broad range of laser exposure conditions, to significantly advance the current pool of knowledge of laser-induced oxidation of chromium. In particular, thin coatings of chromium on a glass substrate were irradiated with nanosecond (ns) and femtosecond (fs) laser beams. The coatings were then characterized to investigate the role of laser parameters in tailoring the morphology, micro-structuring and surface coloration, and depth profiling of the oxidation layer induced. The majority of studies on laser oxidation of chromium films do not report on the depth profiling of the $Cr_2O_3$ film in relation to laser fluence. In this study, we have employed Rutherford Backscattering Spectrometry to follow the evolution of the depth profile of the film after irradiation with femtosecond and nanosecond laser sources to study the effect on the surface and depth on the film due to the two laser sources.

## 2. Experimental methods

### 2.1. Sample preparation

Coatings of chromium metal were deposited on a borosilicate glass substrate using a vacuum coating system, which accelerates hot electrons to evaporate the 99.99% pure chromium target onto a glass substrate. Before the plume of the chromium was released to coat the glass support, the vacuum chamber was evacuated to a base pressure of $1\times10^{-7}$ mbar. 200 nm thick coatings of chromium were then deposited on the substrates. The thickness and the time of deposition of the film was measured using a crystal balance in the chamber.

### 2.2. Laser treatment

A 250-ns pulse duration, 1064-nm central wavelength, Nd:YAG fiber laser (YLP-1/100/50/50 from IPG Photonics, Oxford, MA, USA) and a Yb:KGW (Pharos from Light Conversion, Vilnius, Lithuania) 300-fs pulse duration, 1030-nm central wavelength laser were employed to irradiate the chromium thin films. The general characteristics of both laser sources are shown in Table 1.

**Table 1.** The general characteristics of the laser and the fabrication parameters used for irradiation of chromium thin films.

|  |  | Nanosecond | Femtosecond |
|---|---|---|---|
| **Wavelength** | $\lambda$ | 1064 nm | 1030 nm |
| **Pulse duration** | $\tau$ | 250 ns | 300 fs |
| **Repetition rate** | $R$ | 20 kHz | 200 kHz |
| **Maximum pulse energy** | $E$ | 1 mJ | 1 µJ |
| **Beam quality factor** | $M^2$ | 1.7 | 1.2 |
| **Offset from focus** | $z_{off}$ | 100 µm | 104 µm |
| **Spot diameter** | $d_s$ | 124 µm | 42 µm |

A crucial parameter of the laser exposure is the laser net fluence, which is given by the following equation:

$$\text{NF} = \frac{2\omega R F}{v} \quad (1)$$

where, $\omega$ is the beam waist radius at the sample surface, $R$ is the laser repetition rate, $F$ is the peak fluence per-pulse and $v$ is the scan speed of the translational stage. The net fluence accounts for the per-pulse fluence and the effective number of pulses overlapping within the spot size. The net fluence was varied systematically to study its effect on oxidation.

For nanosecond irradiation, the laser beam was focused onto the sample surface using a 2D galvanometer scan head (TSH 8319 by Sunny Technology, Beijing, China) equipped with an F-Theta lens with 100 mm focal length. For femtosecond irradiation, the laser beam was focused using a 0.4 NA objective (20× Mitutoyo Plan Apo Infinity Corrected Long WD Objective, Kawasaki, Japan) onto the surface of the sample placed on a high resolution 3-axis translational stage (FiberGlide 3D series, Aerotech, Pittsburg, USA). In both cases, the sample was translated away from the focal spot in the focal axis, to achieve a larger spot diameter ($d_s$), reported in Table 1 as the offset from focus, $z_{off}$. 2D squares with side length of 1 cm were laser irradiated on the thin film. For both cases, of laser treatment the spacing between the laser modification lines was 25 µm.

*2.3. Diffraction and electron microscope study*

High resolution scanning electron microscope (Auriga-Ziess) was used for further studies of the coating. The apparatus was configured for the use of an in lens detector, which can process both the secondary electrons and backscattered electrons to generate high quality images. The working distance during the raster scanning of the sample was 5.5 mm from the detector, while the accelerating voltage of the field emission source was 5 kV. X-ray diffraction spectrometry was used to study the microstructure of the chromium coatings. The experiments were performed in a Bruker AXS-D8 Advance θ-2θ X-ray diffractometer using the CuK$_{α1}$ X-ray line with a wavelength of 0.154 nm. The step size of the measurement angle was set to 0.1°/min and the sample was scanned in the range of 10° to 90°.

*2.4 Depth profiling*

Rutherford backscattering spectrometry, employing α - particles was used to study the depth profile of the oxidation layers. The α – particles accelerated with an energy of 1.57 MeV were made to bombard the sample. The geometry of the setup was configured to detect the backscattered α - particles for an angle of incidence of 146º, with the measured beam charge of 30ºC and the current being set at 13 nA. The sample stage mounted on a stepping mechanical ladder allowed the measurement of varies samples to be performed in sequence without having to configure the parameters for each sample.

*2.5. Spectroscopic study*

In order to study the chromaticity of the irradiated samples, a white light source (DH-2000-BAL, OceanOptics) was illuminated on the surface with a 2 mm spot size diameter and the reflected spectrum was collected on UV-Vis-NIR Ocean optics spectrometer (MAYA, OceanOptics, Largo, USA). The average scan of the measurements was set at 5 scans per spectrum and the box card value was set at 2.

## 3. Results and discussion

The thin films of chromium on glass substrate were treated with nanosecond and femtosecond laser. The average power and the scan speeds were varied appropriately so as to keep the laser net fluence comparable for both the ns and fs processing. The laser net fluences used were 15, 30, 60, 70, 80 and 100 J/cm$^2$.

*3.1 Scanning electron microscopy*

The nanosecond and femtosecond laser beams interact differently with the chromium thin films due to a difference in the heat profiles during energy deposition [18,19]. In the case of nanosecond laser beam, initially, the nanosecond beam is absorbed by the electronic gas of the atoms in the coating, thus resulting in a rapid increase of the electronic temperature. This is followed by a heat transfer to the crystal lattice of the coating via electron-phonon coupling. The relaxation time of the electronic temperature is longer with the nanosecond laser as compared to the femtosecond laser interaction [20]. This leads to longer residual heat in the coatings exposed to the nanosecond laser which implies greater heat accumulation in the coatings which can result in the lattice damage, which can be observed on the morphology of the treated sample [21]. It should be noted that the thickness of the films is much smaller than the heat diffusion length with nanosecond pulses, which is in the range of few micrometers. Hence, further heat accumulation is expected compared to bulky coatings.

From the scanning electron microscope (SEM) micrographs in Figure 2, both the nanosecond and femtosecond laser exposed coatings showed modification of their surface morphologies due to laser heating. The chromium coating exposed to nanosecond laser pulses with a net fluence of 15 J/cm$^2$ revealed cracks, which are a sign of increased strain of the heated chromium coating. Chromium coating treated with higher net laser fluence shows an increase in the density of cracks on the surface, which can be attributed to the increase in the deposited heat on the top layers of the coating. This results in the increase of the surface temperature, and when the surface cools, it develops a strain between the top and the bottom layers, due to different cooling rates stemming from the non-isothermal conditions.

In contrast, when the chromium coating was exposed to femtosecond laser pulses, no evidence of cracking was observed for net fluences below 15 J/cm$^2$, with only the localized heat craters are observed in the focal spot. However, at net fluences of 60 J/cm$^2$ and greater, cracks were observed, producing heat marks on the coating along the focused laser beam path. Gedvilas *et al.* [22] observed peeling and cracking from the interaction of the coatings deposited using different techniques (DC-sputtering and e-beam). The results showed that the interaction of the laser pulses with the coatings deposited using electron beam peels and crack easily due to lower adhesion and cohesion strength of the coating due to relatively less energetic atoms impinging the glass support during deposition. This phenomenon could also contribute to the peeling observed at lower fluences, since the coatings were deposited using electron beam. Table 2 outlines the observations and possible reasons for it, for each laser net fluence.

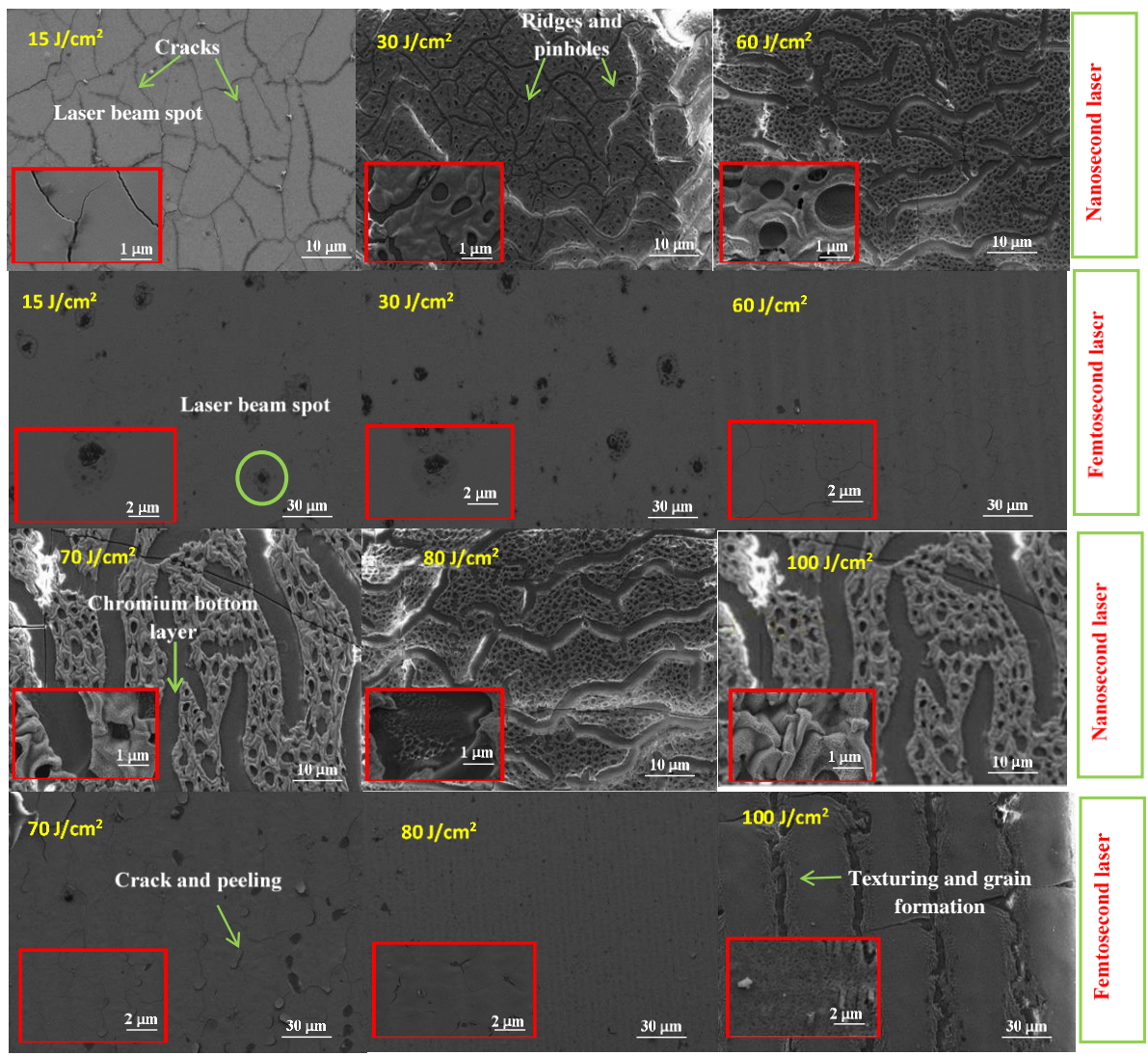

**Figure 2**. SEM images of laser induced modifications on chromium film, with higher magnification micrographs inserted in the figures.

**Table 2.** Scanning electron microscope observations after laser treatment

| Laser source | Net fluence | Observation | Reason |
|---|---|---|---|
| Nanosecond | 15 J/cm$^2$ | Cracks on the surface of the coating | Strain introduced in the coating due to the different cooling rates between the bulk and the upper layer |
| Femtosecond | | Heat marks/spots on the surface of the coating | Localized heating due to the transfer of thermal energy to the coating |
| Nanosecond | 30 J/cm$^2$ | Ridges and pinholes of the surface of the coating | Increase in the internal strain of the coating due to increased temperature, pinholes and porosity are from recrystallized melt |
| Femtosecond | | Heat marks/spots on the surface of the coating | Central crater with melting around, due to the Gaussian nature of the beam |
| Nanosecond | 60 J/cm$^2$ | Wider ridges and bigger pinholes on the surface of the coating | Melt recrystallization from higher temperatures increase the strain, resulting in bigger islands with higher porosity/pinholes sizes |
| Femtosecond | | Cracks and laser heat strips on the surface of the coating | The deposited energy in bulk is non-uniform, resulting in different cooling rates in the bulk resulting in strain induced micro-cracks |
| Nanosecond | 70 J/cm$^2$ | Widen ridges with pinholes that are bigger with texturing/wrinkles around them | Melt recrystallization at elevated temperatures increase strain, increasing porosity of the island |
| Femtosecond | | Micro-cracks and peeling of the coating | The deposited heat is enough to form micro-cracks that can result in peeling of the coating due to higher strain |
| Nanosecond | 80 J/cm$^2$ | Ridges and pinholes of the surface of the coating | Melt recrystallization at elevated temperatures increase strain, increasing porosity of the island |
| Femtosecond | | Texturing of the top surface, micro-crack and peeling | The molten region nucleates and form micro-grains, with the residual heat forming micro-cracks due to strain |
| Nanosecond | 100 J/cm$^2$ | Wrinkles, bigger pinholes, wider ridges, texturing and under-layer of chromium coating | Melt recrystallization at elevated temperatures increase strain, increasing porosity of the island |
| Femtosecond | | Ablation of the coating and ridge formation | Peak of the Gaussian beam where energy is highest, results in ablation, and a texturing is formed around the focal spot |

From the theoretical calculations of the laser-matter interactions and some experimental observations from other research groups [23], it is deduced that at higher fluence there is melting and subsequent recrystallization of the coating. This can be clearly seen in the case of nanosecond laser exposure at higher fluence, wherein the top layer of the coating form islands when the melt recrystallizes. This is not the case for the femtosecond laser exposed coatings since at higher fluence the interaction of the coating with the laser beam results in ablation of the coating, as observed in Figure 2.

*3.2 X-ray diffraction*

All the chromium coatings exposed to ns laser and fs laser were studied for their morphology to quantify the micro structuring induced by laser. The X-rays diffraction patterns of the coatings in Figure 3, shows the evolution of the coatings with respect to the laser fluence. At a lower fluences of 15 J/cm$^2$, chromium diffraction peaks are dominant and there is no emergence of other crystallographic peak due to laser exposure. This applies to both the nanosecond and femtosecond laser irradiation. As the laser fluence increases to 30 J/cm$^2$, there is an appearance of the sharp $Cr_2O_3$ and narrow Bragg peaks at $2\theta = 36.2°$, which corresponds to the (110) rhombohedral crystallographic phase, as observed from the JCPD card 00-038-1479. This evidence implies the formation of $Cr_2O_3$ crystallographic phase. This crystallographic phase is formed on top of the chromium coating, as can be clearly observed from the scanning electron micrographs in Figure 2 showing a textured top layer that is porous and an under layer of chromium which has not been textured. The thickness of the $Cr_2O_3$ layer formed is also related to the penetration depth of laser photons at the infrared wavelength range at a given fluence.

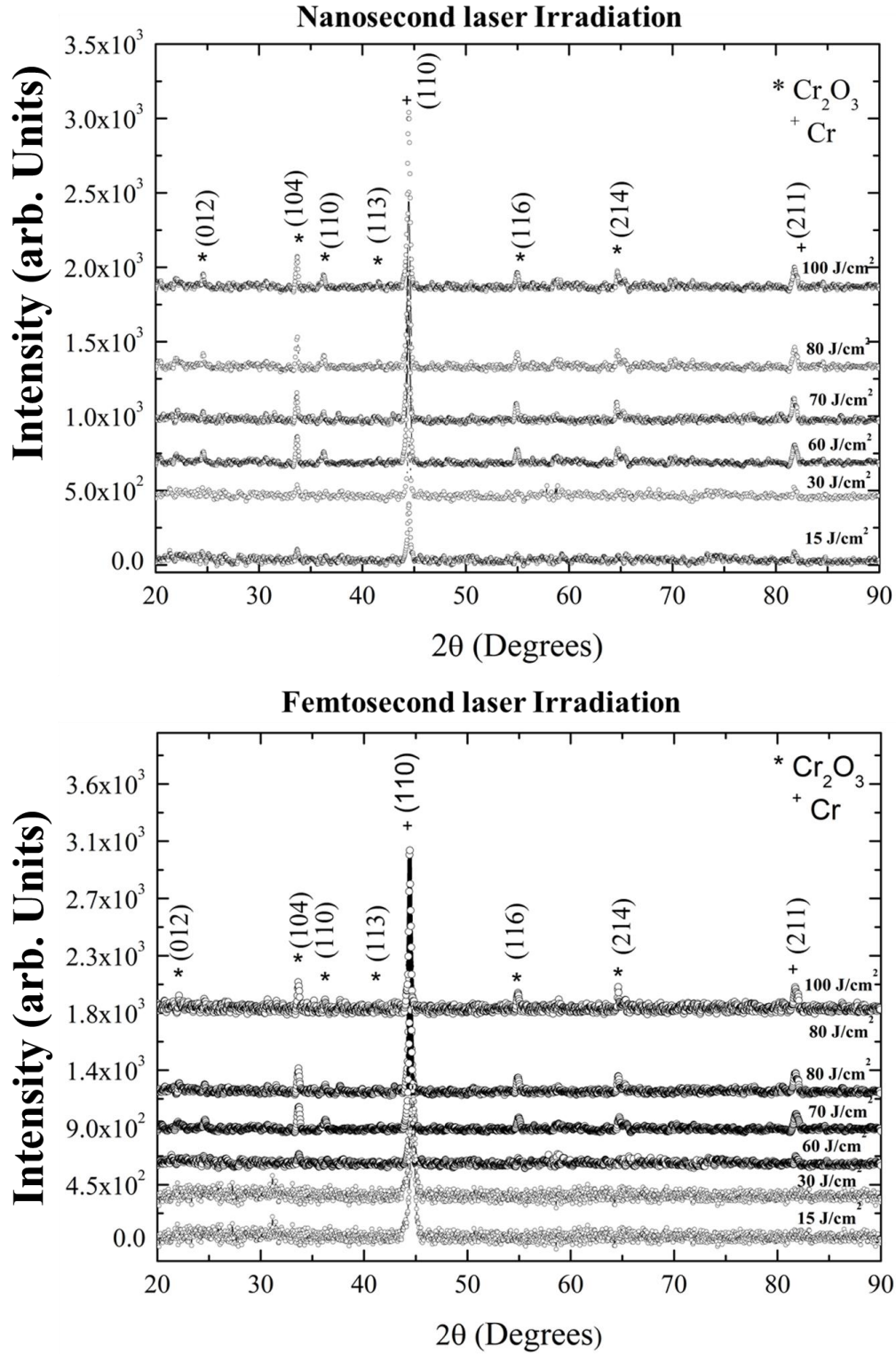

**Figure 3**. X-ray diffraction patterns of chromium coatings treated with nanosecond and femtosecond laser.

Furthermore, it is observed that as the laser net fluence increases, for both ns laser and fs laser (60 J/cm$^2$ to 100 J/cm$^2$) the intensity of the Bragg peaks of $Cr_2O_3$ phase increases, which can be attributed to the increase in the crystal sizes and/or increase in the number of $Cr_2O_3$ molecules formed. From the X-ray diffraction patterns of both the ns and fs laser the trend seems to be that, as the laser net fluence increase, the oxide layer Bragg peaks increase. This phenomenon is accompanied by the emergence of other $Cr_2O_3$ phases (012), (104), (113), (116) and (214) as shown by the Bragg peaks at $2\theta = 24.5°, 33.6°, 41.5°, 58.9°$, respectively.

*3.3 Rutherford backscattering spectrometry - (Depth profiling)*

Figure 4(b) shows the spectra of the Rutherford backscattering spectrometry with normalized yield on the *y*-axis and the channels and energy of the α – particles that have interacted with the coatings on the *x*-axis. From the spectra of the coatings exposed to various laser fluences of ns and fs laser sources, it can be observed from Figure 4(b) that the yield of the chromium normalized peak and the peak width decreases as the laser fluence increases. This can be attributed to two phenomena, (i) non-uniformity of the coating due to cracks formed on the coating resulting in clusters formation, and (ii) the ablation of the coating at higher fluences, resulting in less atoms backscattering the alpha particles which affects the normalized yield. These observations can be related with the scanning electron microscope micrographs in Figure 2, and can be explained as follows, when the α – particle from the accelerated beam interact with the laser treated coatings on the non-uniform portion of the coating due to cracking and/or roughness, the Rutherford backscattered alpha particles suffer multiple interaction with the sample as shown in the schematic in Figure 4(a), resulting in higher energy loss. Thus, Rutherford backscattered alphas particles have lower energy and are assigned to lower channels by the software. It can be seen that, the higher the fluence, the more non-uniform the coating and the chromium peak (approximately 1.2 MeV) of the Rutherford backscattered spectrometry plot shifts.

Figure 4. (c) – (f) shows a representative Rutherford backscattered spectrometry plots of the laser treated coatings as the net fluence increase. Plot (c) is a typical spectrum of the $Cr_2O_3$ spectrum at lower net fluence. At a channel value of 500, it can be observed that the normalized yield shows an edge-like transition, which becomes smoother shouldered as the laser fluence increases. This phenomenon is a result of surface roughness and cracks formed on the coatings, which influence the backscattered alphas, such that the yield is distorted. Furthermore, the top edge of the chromium peak changes from being sharp edge into a smoother curve as seen in (e) and (f). This could be attributed to the migration of the oxygen atomic species in the coating and also a small concentration of the chromium atoms migrating into the glass support. This phenomenon was reported [24,25] when they observed that when there is a migration of atoms from the supporting layer of the coating to the top of the coating. The edge of the spectra tends to change. In our case, it could also be that at higher fluence there is a slight migration of constituent atoms of the glass substrate due to the increased penetration depth of the photons, resulting in heating of the glass support.

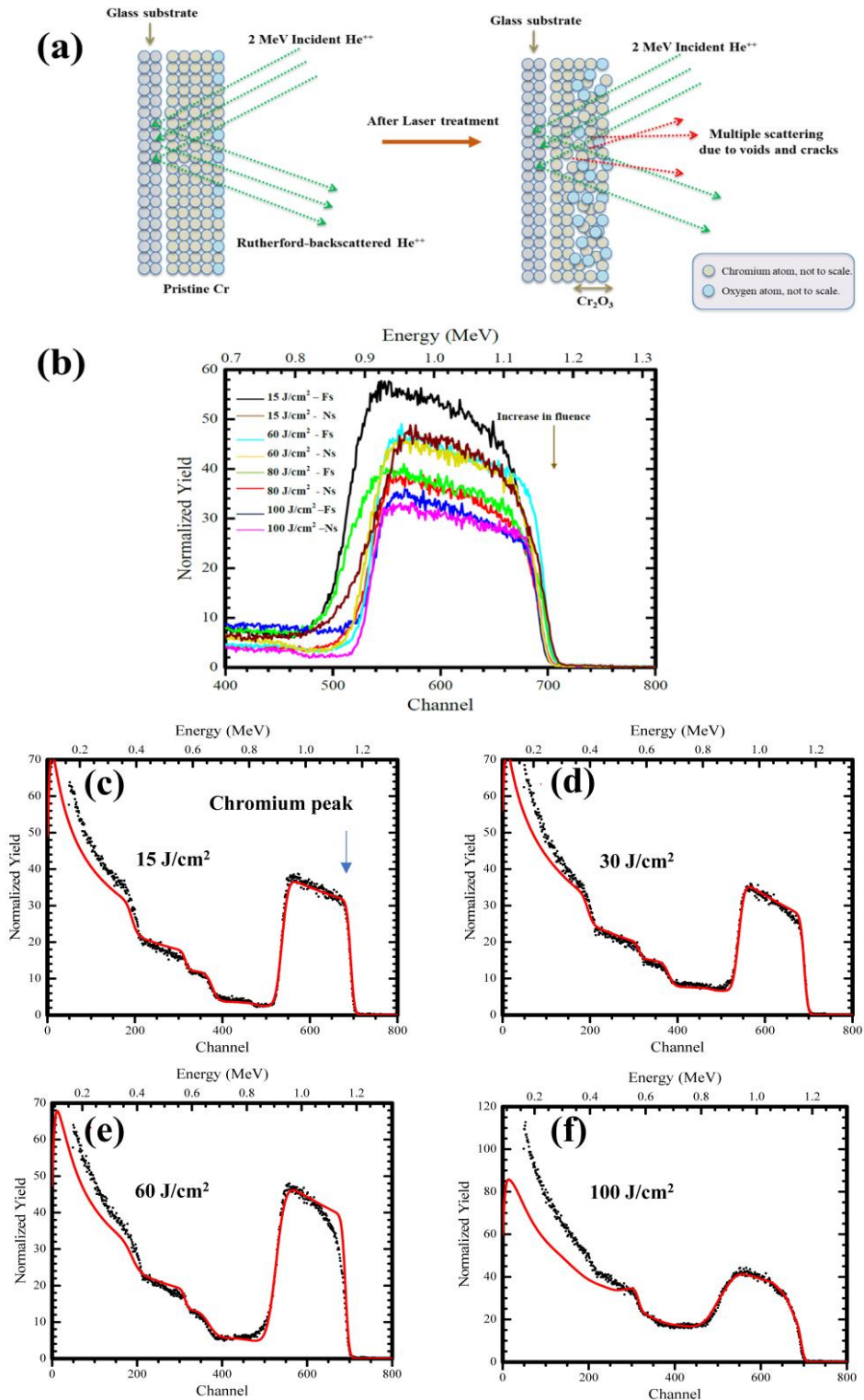

**Figure 4**. (a) Schematic representation of the Rutherford backscattering spectrometry of the coatings. (b) Rutherford backscattering spectrometry plot of the nanosecond and femtosecond laser treated coatings with increasing net fluence. (c) – (f) Chromium peaks from the Rutherford backscattered spectrometry of the coating treated with energetic photons.

From the Rutherford Manipulator Universal Program (RUMP) software, the oxygen concentration in the thin coatings increases as the laser fluence increases. This is due to the incorporation of the oxygen atoms from the ambient while exposing the coating to energetic laser photons. These atoms first sit on the interstitial positions of the chromium lattice. Once there is sufficient energy of formation to bond with chromium atoms, chromium oxide is formed. As the formation energy of other crystallographic phases of $Cr_2O_3$ is made available, those phases are formed with the available oxygen atoms. This can be observed from the x-ray diffraction study, where other crystallographic phases of $Cr_2O_3$ appear at higher fluence.

*3.4 Chromaticity*

The color of the thin coatings after laser treatment changed depending on the laser source and the laser fluence used. The colors varied from light blue to amber and violet hues. The color evolution of the nanosecond and femtosecond laser treated coatings were different. Figure 5 shows the CIE chromaticity diagrams of the coatings as indicated by black dot, representing the color at a given fluence. In the case of the nanosecond laser, the chromaticity change followed no particular discernable trend, whereas in the case of the femtosecond laser treated coatings, the hue of the coatings was mostly in the amber region of the CIE chromaticity diagram, and only at the highest fluence of 100 J/cm², a significant change in the hue was observed.

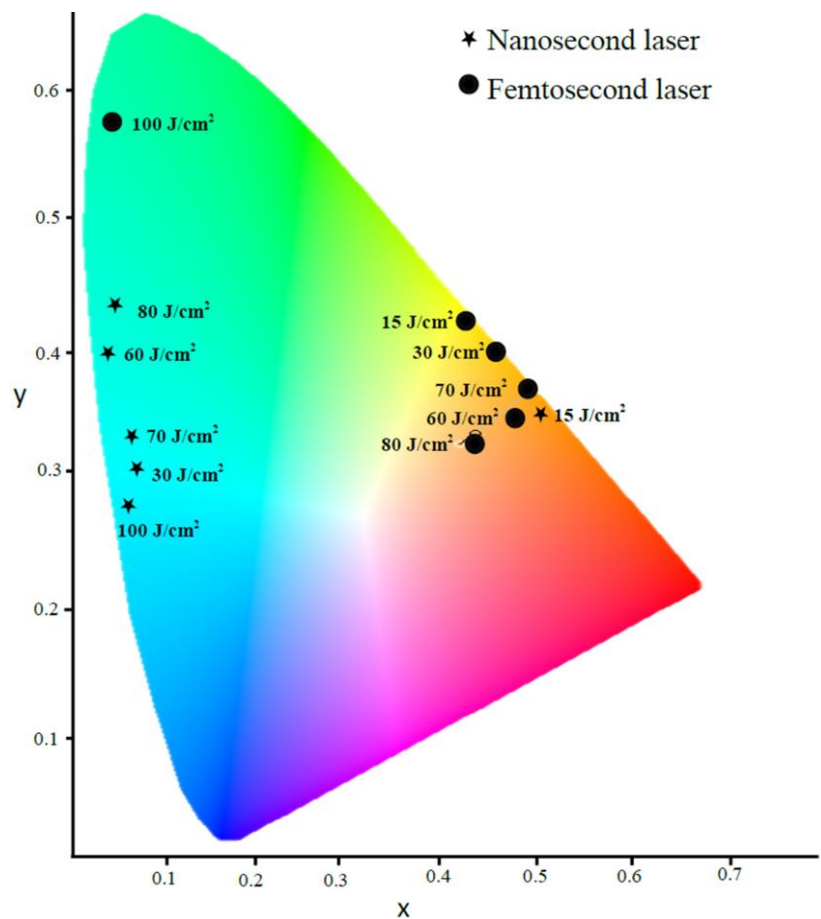

**Figure 5.** CIE Chromaticity diagram of the nanosecond and femtosecond laser treated chromium coatings.

The colors of a hue of the nanosecond laser treated samples and the color of the hue of the femtosecond laser treated samples, showed to have no correlation to microstructure, but the difference in chromaticity between the nanosecond and the femtosecond treated sample, could be related to the way light is diffusely reflected due to the morphology thin coatings of nanosecond and femtosecond treatment. This color change of the coating after the exposure to the energetic photons from various laser source have been observed when metals are laser treated in ambient conditions. The colors are usually attributed to plasmonic effect of the coating and the diffraction of light by the nano-crystals formed due to laser interaction [26].

## 4. Conclusions

From the study of the microstructure, morphology and surface depth profiling (using X-ray diffraction, SEM and RBS, respectively) on the chromium thin coatings exposed to femtosecond and nanosecond laser, it was observed that the crystallographic phases of the chromium oxide formed was same for both cases. Further probing of these lasers treated coatings using scanning electron microscope, revealed that the morphology of the sample differs significantly, this was attributed to the distinctly different nature of interaction of nanosecond laser with chromium coatings compared to the femtosecond laser. Chromaticity results showed a differed response for nano and femtosecond irradiation, which could be due to microstructural differences between the nanosecond laser treated samples and femtosecond. The depth profiling of the films using Rutherford backscattering spectrometry confirmed the migration of oxygen in the films and ultimate formation of an oxide layer. From the results it can be inferred that the $Cr_2O_3$ formation in thin films using the nanosecond and the femtosecond laser follows similar oxygen diffusion as for forming $Cr_2O_3$ chemically.

**Acknowledgments**

The author would like to express a sincere gratitude to the National Research Foundation of South Africa and the UNESCO AFRICA CHAIR IN NANOSCIENCE AND NANOTECHNOLOGY. We thank Professor Guglielmo Lanzani and Dr Luigino Criante for the use of the FemtoFab facility at CNST - IIT Milano for the laser fabrication experiments.